\documentclass[12pt]{article}

\usepackage{amssymb}
\usepackage{amsmath}
\usepackage{amsthm}
\newtheorem{thm}{Theorem}

\theoremstyle{definition}

\newtheorem{obv}[thm]{Observation}
\newtheorem*{bproof}{Proof}
\theoremstyle{remark}

\newcommand {\N}{\mathbb{N}}
\newcommand {\Z}{\mathbb{Z}}

\begin{document}



\begin{center}
{\Large \bf The Frequent Paucity of Trivial Strings}
\end{center}


\begin{center}
{{\large Jack H. Lutz}\\
Department of Computer Science\\
Iowa State University\\
Ames, IA 50011, USA\\
lutz@cs.iastate.edu
}
\end{center}

\begin{abstract}
A 1976 theorem of Chaitin can be used to show that arbitrarily dense sets of lengths $n$ have a {\em paucity of trivial strings} (only a bounded number of strings of length $n$ having trivially low plain Kolmogorov complexities).  We use the probabilistic method to give a new proof of this fact.  This proof is much simpler than previously published proofs, and it gives a tighter paucity bound.
\end{abstract}

% \linenumbers

\section{Background}
\label{S1}
A string of binary data is {\em trivial} if, like a string of all zeros, it contains negligible information beyond that implicit in its length.  This notion of triviality has been made precise in several different ways, and these have been useful in the foundations of Kolmogorov complexity \cite{jLove69}, information-theoretic characterizations of decidability and polynomial-time decidability \cite{jChai76,jOKSW94}, formal language theory \cite{jLiVit95}, and the theory of K-trivial sequences \cite{bNies09,bDowHir10}.

These applications share several common features.  Each uses some version of Kolmogorov complexity to quantify the information content of a string.  Each {\em parametrizes} its triviality notion by a nonnegative integer $c$, defining a string to be $c$-{\em trivial} if its information content is within $c$ bits of a triviality criterion. Most crucially, the key to each of these applications is a {\em paucity theorem}, stating that there are many lengths $n$ at which there is a paucity (at most a fixed multiple of $2^c$) of $c$-trivial strings of length $n$.

The first such paucity theorem, reported in 1969, was proved by Meyer \cite{jLove69}.  Chaitin subsequently strengthened Meyer's proof, slightly relaxing his triviality notion and obtaining the following.
\begin{thm}[Chaitin \cite{jChai76}] \label{pautheorem} There is a constant $a \in \N$ such that, for all $n, d \in \N$, at most $2^{d+a}$ strings $x \in \{0, 1\}^n$ satisfy $C(x) \leq d + C(n)$.
\end{thm}

Here $C(x)$ is the {\em plain Kolmogorov complexity} of $x$, the minimum number of bits required to program a fixed universal Turing machine to print the string $x$, and $C(n) = C(s_n)$, where $s_0, s_1, \ldots$ is a standard enumeration of $\{0, 1\}^\ast$.  (Thorough treatments of $C(x)$ appear in \cite{bLiVit08,bNies09,bDowHir10}.)

This note concerns paucity theorems involving $\log n$, rather than $C(n)$, as a triviality criterion.  Since $C(n)$ is usually close to $\log n$, one such paucity theorem can be derived from Theorem~\ref{pautheorem}, as we now show.

Logarithms here are base-2.  We will use the ({\it Schnirelmann}) {\em density} of a set $L \subseteq \N$, which is
\[
\sigma(L) = \inf \left\{ \left.\frac{|L_{<m}|}{m} \; \right| \;m \in \Z^+\right\},
\]
where we write $L_{<m} = L \cap \{0, \ldots , m-1 \}$ \cite{bTaoVu06}.  Intuitively the condition $n \in L$ holds {\em frequently} if $\sigma(L) > 0$.  This is clearly a stronger condition than the assertion that $L$ is infinite.  To relate the triviality criteria $\log n$ and $C(n)$, define the set
\[
L(r) = \left\{ n \in \N  \left. \rule{0pt}{14pt} \right| \; C(n) + r \geq \log n \right\}
\]
for each $r \in \N$.

\begin{obv}
For each $R \in \N$, $\sigma(L(r)) \geq 1 - 2^{1-r}$.
\end{obv}

\begin{bproof}
For each $m \in \Z^{+}$, the complement $L(r)^c$ of $L(r)$ satisfies
\begin{eqnarray*}
           \nonumber (L(r)^c)_{<m} &=& \left.\left\{ n < m \;\right| C(n) < (\log n) - r \right\} \\
            &\subseteq &  \left.\left\{ n < m \;\right| C(n) < (\log m) - r \right\}
         \end{eqnarray*}
so
\begin{eqnarray*}
  \nonumber \left|(L(r)^c)_{<m}\right| &\leq& \left|\{0,1\}^{<(\log m) - r}\right| \\
   &<& 2^{1-r+\log m} \\
   &=& 2^{1-r}m.
\end{eqnarray*}
It follows that
\begin{eqnarray*}
  \nonumber \sigma(L(r)) &=& \inf\left\{\left. \frac{|L(r)_{<m}|}{m} \;\right| m \in \Z^{+}\right\} \\
   &\geq& \inf \left\{ \left. \frac{m - 2^{1-r} m}{m} \;\right| m \in \Z^{+} \right\} \\
   &=& 1 - 2^{1-r}.
\end{eqnarray*}
\hfill $\Box$
\end{bproof}

We now have the following easy consequence of Theorem~\ref{pautheorem}.
\begin{thm}[very frequent paucity theorem] \label{vfpt} The constant $a$ of Theorem~\ref{pautheorem} has the property that, for all $c, r \in \N$, the set of nonnegative integers $n$ for which at most $2^{c+a+r}$ strings $x \in \{0, 1\}^n$ satisfy $C(x) \leq c + \log n$ has density at least $1 - 2^{1-r}$.
\end{thm}

\begin{bproof}
Let $a \in \N$ be as in Theorem~\ref{pautheorem}, and let $c, r \in \N$.  For each $n \in \N$, define the sets
\[
B_n = \left\{ \left. x \in \{0, 1\}^n \right| C(x) \leq c + \log n \right\}
\]
and
\[
B^{\prime}_n = \left\{ \left. x \in \{0, 1\}^n \right| C(x) \leq c + r + C(n) \right\}
\]
and let
\[
L_c = \left\{ n \in \N \left| |B_n| \leq 2^{c+a+r} \right\} \right..
\]
It suffice to show that $\sigma(L_c) \geq 1 - 2^{1-r}$.

Let $n \in L(r)$.  Then $C(n)+r \geq \log n$, so $B_n \subseteq B^{\prime}_n$.  Applying Theorem~\ref{pautheorem} with $d = c+ r$, we have $|B^{\prime}| \leq 2^{c+r+a}$, whence $|B_n| \leq 2^{c+ r + a}$.  Hence $n \in L_c$.

We have now shown that $L(r) \subseteq L_c$.  It follows by Observation~2 that
\[
\sigma(L_c) \geq \sigma(L(r)) \geq 1 - 2^{1-r}.
\]
\hfill $\Box$
\end{bproof}

The proofs of Theorem~\ref{pautheorem} and Meyer's earlier paucity theorem are somewhat involved.  Part of this is because these early proofs were aimed at proving more, namely that
\begin{enumerate}
\item[(I)] for every $c \in \N$ there are at most $2^{c+a}$ {\em infinite} binary sequences that are $c$-{\em trivial} in the sense that {\em every} nonempty prefix $x$ of such a sequence satisfies $C(x) \leq c + log |x|$; and
\item[(II)] every such $c$-trivial sequence is decidable.
\end{enumerate}
It is clear that (I) follows immediately from Theorem~\ref{pautheorem}, and it is now well understood that (II) follows directly from (I), because every isolated infinite branch of a decidable tree is decidable \cite{bDowHir10}.

In the 1990s, Li and Vitanyi proved the following paucity theorem.

\begin{thm}[Li and Vitanyi \cite{jLiVit95}]
There is a constant $a \in \N$ such that, for every $c \in \N$, there exist infinitely many lengths $n$ for which at most $2^{c+a}$ strings $x \in \{0, 1\}^n$ satisfy $C(x) \leq c + \log n$.
\end{thm}

Theorem~4 is weaker than Theorem~3, because it only tells us that the paucity of trivial strings occurs at infinitely many lengths.  Li and Vitanyi's proof of Theorem~4 is simpler than the proof of Theorem~1 (hence simplier than the proof of Theorem~3), even when one discounts the parts of the proof of Theorem~1 devoted to (I) and (II).  However, even Li and Vitanyi's simplified proof is nontrivial.

\section{Result}

The purpose of this note is to give a {\em very} simple proof of a frequent paucity theorem.  Our theorem's frequency condition is as strong as that of Theorem~3.  However, our theorem improves on earlier paucity theorems in a significant respect:  While the proofs of Theorems 1, 3, and 4 require the constant $a$ to be as large as the number of bits required to encode a nontrivial Turing machine, our simple proof shows that it suffices to take $a=1$.

Our simple proof has a simple intuition: As in the proof of Theorem~3, let
\[
B_n = \left\{ \left. x \in \{0, 1\}^n \right| C(x) \leq c + \log n \right\}.
\]
We want to show that $|B_n|$ is often small.  Well, the {\em average} of the first $m$ values of $|B_n|$ is
\begin{eqnarray*}
  \nonumber \frac{1}{m} \sum_{n=0}^{m-1}|B_n| &=& \frac{1}{m} \left| \bigcup_{n=0}^{m-1} B_n \right| \\
   &\leq& \frac{1}{m} \left| \{0, 1\}^{< c + \log m} \right| \\
   &<& \frac{1}{m} 2^{c+\log m + 1} \\
   &=& 2^{c+1},
\end{eqnarray*}
so $|B_n| \leq 2^{c+1}$ must hold frequently!  The details follow.

\begin{thm}
Let $c \in \N$.

{\em 1  (frequent paucity).}
The set of nonnegative integers $n$ for which at most $2^{c+1}$ strings $x \in \{0, 1\}^n$ satisfy $C(x) \leq c + \log n$ has density at least $(2^{c+1} - 1)^{-1}$.

{\em 2 (very frequent paucity).}
For every $r \in \N$, the set of nonnegative integers $n$ for which at most $2^{c+r}$ strings $x \in \{0, 1\}^n$ satisfy $C(x) \leq c + \log n$ has density at least $1 - 2^{1-r}$.
\end{thm}

\begin{bproof}
Let $c, r \in \N$, and let $d = 2^{c+r}$.  For each $n \in \N$, let
\[
B_n = \left\{ \left. x \in \{0, 1\}^n \right| C(x) \leq c + \log n \right\},
\]
noting that $B_0 = \emptyset$, and let
\[
L = \left\{ n \in \N \left|\rule{0pt}{14pt}\right. |B_n| \leq d \right\}
\]
Let $m \in \Z^{+}$, and let $l = |L_{<m}|$.

Consider the average
\[
\mu = \frac{1}{m} \sum_{n=0}^{m-1} | B_n|.
\]
We have
\begin{eqnarray*}
  \nonumber \mu &=& \frac{1}{m} \left| \bigcup_{n=0}^{m-1} B_n \right| \\
   &\leq& \frac{1}{m} \left| \{0, \}^{<c | log m} \right| \\
   &<& \frac{1}{m}2^{c+ \log m + 1} \\
   &=& 2^{c+1}
\end{eqnarray*}
and
\[
\mu \geq \frac{1}{m} (m - l)(d+1)
\]
whence
\begin{equation*} \tag{$\ast$}
m \cdot 2^{c+1} > (m-l)(d+1).
\end{equation*}

1. If $r=1$, then ($\ast$) says that
\[
md > (m-l)(d+1)
\]
whence
\[
l > \frac{m}{d+1}.
\]
Since this holds for all $m \in \Z^{+}$, it follows that $\sigma(L) \geq \frac{1}{d+1}$

\vspace*{6pt}
2. More generally, for $r \in \N$, ($\ast$) implies that
\[
m \cdot 2^{c+1} > (m - l)2^{c+r},
\]
whence
\[
d > m(1 - 2^{1-r})
\]
Since this holds for all $m \in \Z^{+}$, it follows that $\sigma(L) \geq 1 - 2^{1-r}$. \hfill $\Box$
\end{bproof}

\section{Conclusion}
The simplicity of the above proof is the main contribution of this note.  Its simplicity arises from its use of the first moment probabilistic method \cite{bAloSpe08,bTaoVu06}: Rather than deal with the cardinalities $|B_n|$ individually, it examines their average.  It is an open question whether the probabilistic method can similarly simplify the proof of Theorem~1.

A brief remark on pedagogy: Li and Vitanyi's Kolmogorov complexity characterization of regular languages \cite{jLiVit95,bLiVit08} yields a simple and intuitive method for proving that languages are not regular.  A possible obstacle to teaching this method in undergraduate theory courses has been that the characterization theorem relies on the (seemingly) difficult Theorem~4.  The simple proof here removes that obstacle.

\section*{Acknowledgments}
I thank the referees for extremely useful observations.
This research was supported in part by National Science Foundation Grant 0652569.  Part of this work was done during a sabbatical at Caltech and the Isaac Newton Institute for Mathematical Sciences at the University of Cambridge.


\bibliographystyle{plain}
\bibliography{master}


\end{document}